\documentclass[prd,eqsecnum,twocolumn,showpacs,amsmath,amssymb]{revtex4}

\usepackage{graphicx}

\usepackage{bm}

\newcommand{\beq}{\begin{equation}}
\newcommand{\eeq}{\end{equation}}
\newcommand{\beqs}{\begin{eqnarray}}
\newcommand{\eeqs}{\end{eqnarray}}

\begin{document}

\title{Ground State Entropy of the Potts Antiferromagnet on Homeomorphic
Expansions of Kagom\'e Lattice Strips}

\author{Robert Shrock and Yan Xu}

\affiliation{ C. N. Yang Institute for Theoretical Physics \\
Stony Brook University \\
Stony Brook, NY 11794}

\begin{abstract}

We present exact calculations of the chromatic polynomial and resultant ground
state entropy of the $q$-state Potts antiferromagnet on lattice strips that are
homeomorphic expansions of a strip of the kagom\'e lattice.  The dependence of
the ground state entropy on the form of homeomorphic expansion is elucidated.

\end{abstract}

\pacs{05.50.+q,64.60.Cn,75.10.Hk}

\maketitle

\section{Introduction}

   Nonzero ground state entropy (per lattice site), $S_0 \ne 0$, is an
important subject in statistical mechanics, as an exception to the third law of
thermodynamics and a phenomenon involving large disorder even at zero
temperature. Since $S_0 = k_B \ln W$, where $W = \lim_{n \to \infty}
W_{tot.}^{1/n}$ and $n$ denotes the number of lattice sites, $S_0 \ne 0$ is
equivalent to $W > 1$, i.e., a total ground state degeneracy $W_{tot.}$ that
grows exponentially rapidly as a function of $n$. One physical example is
provided by H$_2$O ice, for which the residual entropy per site (at 1
atm. pressure) is $S_0 = (0.41 \pm 0.03)k_B$, or equivalently, $W=1.51 \pm
0.05$ \cite{pauling35} (a recent study is \cite{berg07}).  In ice, the ground
state entropy occurs without frustration; that is, each of the ground state
configurations of the hydrogen atoms on the hydrogen bonds between water
molecules minimizes the internal energy of the crystal. This is in contrast to
systems where nonzero ground state entropy is associated with frustration,
including the Ising antiferromagnet on the triangular lattice and spin glasses.

  A model that exhibits ground state entropy without frustration and hence
provides a useful framework in which to study this phenomenon is the $q$-state
Potts antiferromagnet (PAF) \cite{wurev}-\cite{chowwu} on a given lattice
$\Lambda$ or, more generally, a graph $G$, for sufficiently large $q$.  An
interesting question concerns how this ground state entropy, or equivalently,
the ground state degeneracy per site, $W$, depends on properties of the
graph. One can study this using such methods as Monte Carlo simulations,
calculations of rigorous upper and lower bounds, and large-$q$ series.  One can
also gain considerable insight from exact solutions for $W$ on $n \to \infty$
limits of certain families of graphs.

 A particular question is how $W$ changes when one inserts new vertices on
certain bonds of the graph.  In mathematical graph theory, this insertion
process is called a homeomorphic expansion of the graph (and the opposite
process, removing degree-2 vertices from bonds of a graph, is called a
homeomorphic reduction). It is useful to answer this question in simple cases
such as lattice strips, since one can get exact explicit analytic results for
these cases \cite{hs,pg}.  In this paper we shall continue this line of study,
extending the results of earlier work that one of us did with S.-H. Tsai
\cite{hs,pg}.  We shall calculate exact expressions for the chromatic
polynomial and resultant ground state degeneracy per site of the $q$-state
Potts antiferromagnet on lattice strips that are homeomorphic expansions of a
strip graph of the kagom\'e lattice.  Our results and their comparison with
analogous exact calculations for the kagom\'e strips without homemorphic
expansion in \cite{strip}-\cite{wcy} and with homeomorphic expansions of
square-lattice ladder graphs in \cite{pg} add to our understanding of the
effect of homeomorphic expansions on the per-site ground state degeneracy and
entropy of the Potts antiferromagnet.

\section{Generalities and Connection with Chromatic Polynomials} 

Let us consider a graph $G=(V,E)$, defined by its vertex (site) set $V$ and its
edge (bond) set $E$.  The number of vertices of $G$ is denoted $n(G) = |V|
\equiv n$, as above, and the number of edges of $G$ is denoted $e(G)=|E|$.  We
use the symbol $\{G\}$ for the limit $\lim_{n(G) \to \infty} G$ of a given
family of graphs, such as the infinite-length limit of a strip graph.  The
$q$-state Potts model partition function at a temperature $T=1/(k_B\beta)$ on
the graph $G$ is $Z(G,q,T) = \sum_{\{\sigma_i\}}e^{-\beta {\cal H}}$, with
Hamiltonian ${\cal H} = -J\sum_{e_{ij}}\delta_{\sigma_i \sigma_j}$, where $J$
is the spin-spin interaction constant, $i$ and $j$ denote vertices on $G$,
$e_{ij}$ is the edge connecting them, and $\sigma_i$ are classical spins taking
on values in the set $\{1,...,q\}$. For the Potts antiferromagnet, $J < 0$ so
that, as $T \to 0$, $\beta J = -\infty$; hence, in this limit, the only
contributions to the partition function are from spin configurations in which
adjacent spins have different values.  The resultant $T=0$ PAF partition
function is therefore precisely the chromatic polynomial $P(G,q)$ of the graph
$G$:
\beq
Z(G,q,T=0)_{PAF} = P(G,q) \ , 
\label{zprel}
\eeq
where $P(G,q)$ counts the number of ways of assigning $q$ colors to the
vertices of $G$ subject to the condition that no two adjacent vertices have the
same color (reviews include \cite{biggs}-\cite{newton}).  This is called a
proper $q$-coloring of (the vertices of) $G$.  Thus, 
\beq
W(\{G\},q) = \lim_{n \to \infty}P(G,q)^{1/n} \ .
\label{w}
\eeq
The determination of $W(\{G\},q)$ is thus equivalent to the determination of
$S_0(\{G\},q)$, and we shall generally give results in terms of $W(\{G\},q)$.
The minimal integer value of $q$ for which one can carry out a proper
$q$-coloring of $G$ is the chromatic number, $\chi(G)$.  In general, for
certain special values of $q$, denoted $q_s$, one has the following
noncommutativity of limits \cite{w}
\beq
\lim_{n \to \infty} \lim_{q \to q_s} P(G,q)^{1/n} \ne
\lim_{q \to q_s} \lim_{n \to \infty} P(G,q)^{1/n} \ ,
\label{wnoncom}
\eeq
and hence it is necessary to specify which order of limits one takes in
defining $W(\{G\},q)$.  Here, by $W(\{G\},q)$ we mean the function obtained by
setting $q$ to the given value first and then taking $n \to \infty$.  For the
families of graphs considered here, the set $\{q_s\} = \{0,1,2\}$.  The
noncommutativity (\ref{wnoncom}) will not be important for our discussion,
since we will restrict our calculations of $W(\{G\},q)$ to $q \ge 3$.  For
lattice strips that are $m$-fold repetitions of some basic subgraph, one can
take a limit $n \to \infty$ by taking the limit $m \to \infty$.

The family of homeomorphically expanded graphs of the kagom\'e lattice strip
that we consider are denoted $[H_k(kag)]_{m,BC}$, where $H$, $kag$, and $BC$
stand for homeomorphic expansion, kagom\'e, and longitudinal boundary
conditions, free (f) or cyclic (c).  A member of this family is defined as
follows.  We start with a minimal-width kagom\'e strip graph, a portion of
which is shown in Fig. 1(f) of Ref. \cite{strip}, comprised of $m$ subgraphs,
each of which consists of a hexagon with its two adjoining triangles.  We then
insert $k$ vertices on each longitudinal edge of a hexagon in this original
kagom\'e strip graph. Thus, the graph $[H_k(kag)]_{m,BC}$ is a strip 
of $m$ subgraphs each of which consists of two triangles and a $p$-gon with
\beq
p = 6+2k \ . 
\label{p}
\eeq
The graph $[H_1(kag)]_{m,BC}$ involves subgraphs with two triangles and
an octagon, and so forth for higher values of $k$. The kagom\'e strip itself is
the case $k=0$.  The chromatic number of the free and cyclic
kagom\'e strips is $\chi=3$, and this remains true for the homeomorphic
expansions $[H_k(kag)]_{m,BC}$:
\beq
\chi([H_k(kag)]_{m,f})=\chi([H_k(kag)]_{m,c})=3 \ . 
\label{chi}
\eeq
We shall sometimes use the abbreviations $kag_{k,m,BC} \equiv
[H_k(kag)]_{m,BC}$ with $BC=f$ or $BC=c$ and, for the family as a whole,
suppressing the $m$ index, $kag_{k,BC} \equiv [H_k(kag)]_{BC}$. 

For the relevant range, $q \ge 3$, of interest here, the $W(\{G\},q)$ functions
computed via the infinite-length limits of the $[H_k(kag)]_{m,BC}$ strips with
free and cyclic (and M\"obius) longitudinal boundary conditions (BC)
are all the same.  Since the calculation is easiest if one uses strip graphs
with free longitudinal boundary conditions, we shall do this.  It is also of
interest to calculate the chromatic polynomials for the corresponding strip
graphs with cyclic boundary conditions and we will do this.  The $m \to
\infty$ limits for these families of homeomorphically expanded kagom\'e strips
will be denoted $\{[H_k(kag)]_{BC}\}$ and, for the $W$ function, which is
independent of the boundary conditions, $W(\{H_k(kag)\},q)$.

As noted above, our exact results for the infinite-length homeomorphically
expanded kagom\'e strip graphs complement other methods of
studying $W$ functions on lattices, such as rigorous bounds, large-$q$ series,
and Monte Carlo measurements \cite{biggs77,w3,wn,wsc}.  Other homeomorphic
expansions of this kagom\'e strip graph are also of interest, e.g., expansions
in which additional vertices are added to edges of the triangles, but here we
shall restrict ourselves to studying the specific homeomorphic expansion
defined above.  In passing, we mention that chromatic polynomials of
homeomorphic expansions of other types of graphs have been studied 
in, e.g., Refs. \cite{hs}, \cite{dktbook}, \cite{wz84}-\cite{peng04}.

\section{Calculational Method}

The chromatic polynomial $P(G,q)$ can be calculated in several ways.  One is
via the deletion-contraction relation.  For a graph $G$, let us denote $G-e$ as
the graph obtained by deleting the edge $e$ and $G/e$ as the graph obtained by
deleting the edge $e$ and identifying the two vertices that were connected by
this edge of $G$.  The latter operation is called a contraction of $G$ on $e$.
Then the chromatic polynomial satisfies the deletion-contraction relation
\beq
P(G,q) = P(G-e,q)-P(G/e,q) \ . 
\label{dcr}
\eeq
$P(G,q)$ can also be determined via the cluster formula \cite{fk} 
\beq
P(G,q) = \sum_{G' \subseteq G} q^{\kappa(G')} \, (-1)^{e(G')} \ , 
\label{fk}
\eeq
where $G'$ is a spanning subgraph, $G'=(V,E')$ with $E' \subseteq E$ and
$\kappa(G')$ denotes the number of connected components in $G'$.  

The numbers of vertices and edges on the $[H_k(kag)]_{m,f}$ and 
$[H_k(kag)]_{m,c}$ graphs are 
\beq
n([H_k(kag)]_{m,c}) = n([H_k(kag)]_{m,f})-3 = (5+2k)m
\label{nkag}
\eeq
and
\beq
e([H_k(kag)]_{m,c}) = e([H_k(kag)]_{_m,f})-2 = (8+2k)m \ . 
\label{ekagk}
\eeq
(For the cyclic strip with $m=1$ some of these are double edges; this does not
affect the chromatic polynomial.)  The graph $[H_k(kag)]_{m,c}$ has vertices
with degrees 3, 4, and, for $k \ge 1$, also 2.  For reference, the infinite 2D
kagom\'e lattice has vertices of uniform degree 4.  Defining, as in
Ref. \cite{wn}, an effective vertex degree, 
\beq
\Delta_{eff} \equiv \lim_{n \to \infty} \frac{2e(G)}{n(G)} \ , 
\label{deltaeff}
\eeq
we have
\beq
\Delta_{eff} = \frac{4(4+k)}{5+2k} \quad {\rm for} \ \{H_k(kag)\} \ . 
\label{deltaeffkagk}
\eeq

Because $\chi([H_k(kag)]_{m,f})=3$, it follows that $P([H_k(kag)]_{m,BC},q)=0$
for $q=0, \ 1, \ 2$ for free or cyclic BC.  Since $P([H_k(kag)]_{m,BC},q)$ is a
polynomial, this implies that
\beq
P([H_k(kag)]_{m,BC},q) \quad {\rm contains \ the \ factor} \quad q(q-1)(q-2) \
. 
\label{pfactors}
\eeq

\section{Strips with Free Longitudinal Boundary Conditions} 

For the family $[H_k(kag)]_f$ of strip graphs $[H_k(kag)]_{m,f}$, it is
convenient to use a generating function formalism, as before
\cite{strip,hs,wa2}.  For arbitrary $k$, this generating function is
\beq
\Gamma([H_k(kag)]_f,q,x) = \sum_{m=0}^{\infty}P([H_k(kag)]_{m+1,f},q)x^{m} \ . 
\label{gamma}
\eeq
The generating function is a rational function in $x$ and $q$ of the form 
\beq
\Gamma([H_k(kag)]_f,q,x) = \frac{a_{k,0} + a_{k,1}x}{1+b_{k,1}x+b_{k,2}x^2} \
. 
\label{gammagen}
\eeq
We write the denominator as 
\beq
1+b_{k,1}x+b_{k,2}x^2 = \prod_{j=1}^2 (1-\lambda_{kag_k,0,j} \, x) \ . 
\label{den}
\eeq

By means of an iterative use of the deletion-contraction relation and 
induction on the homeomorphic expansion parameter $k$, we have
calculated $\Gamma([H_k(kag)]_f,q,x)$ and hence $P([H_k(kag)]_{m,f},q)$ for
arbitrary $k$ and $q$.  Recall that the chromatic polynomial of the circuit 
graph $C_n$ is $P(C_n,q) = (q-1)^n+(q-1)(-1)^n$.  Since this has a factor 
$q(q-1)$, it is convenient to define 
\beq
D_n = \frac{P(C_n,q)}{q(q-1)} = 
\sum_{s=0}^{n-2}(-1)^s {{n-1}\choose {s}} q^{n-2-s}
\label{dk}
\eeq
so that $D_2=1$, $D_3=q-2$, $D_4=q^2-3q+3$, etc. (Where it appears, we shall
write $D_3$ simply as $q-2$.)  
We find (with $p=6+2k$ as given in Eq. (\ref{p})) 
\beq
a_{k,0} = q(q-1)(q-2)^2D_p \ , 
\label{ak0}
\eeq
\beq
a_{k,1} = -q(q-1)^{5+2k}(q-2)^3 \ , 
\label{ak2}
\eeq
\beq
b_{k,1} = -(q-2)(D_p-D_{p-1}+1) \ , 
\label{bk1}
\eeq
and
\beq
b_{k,2} = (q-1)^{3+2k}(q-2)^3 \ . 
\label{bk2}
\eeq
It is readily checked that for the special case $k=0$, these results reduce to
the generating function for the kagom\'e strip given in Ref. \cite{strip}.

Substituting the results for $b_{k,1}$ and $b_{k,2}$ in Eq. (\ref{den}) and
solving for $\lambda_{kag_k,0,j}$, we find
\beq
\lambda_{kag_k,0,j} = \frac{1}{2}(q-2)( D_p-D_{p-1}+1 \pm \sqrt{R_{kkd0}}  \ ) 
\ , 
\label{lamkagkd0}
\eeq
where $p=6+2k$ as given in Eq. (\ref{p}), $j=1,2$ correspond to the $\pm$ 
signs, and
\beq
R_{kkd0} = (D_p-D_{p-1}+1)^2-4(q-1)^{3+2k}(q-2) \ . 
\label{rkagkd0}
\eeq

Using the general methods of \cite{hs} for expressing the chromatic polynomial
in terms of the coefficients in the generating function, we find that 
$P([H_k(kag)]_{m,f},q)$ is given by 
\begin{widetext}
\beq
P([H_k(kag)]_{m,f},q) = \frac{(a_{k,0}\lambda_{kag_k,0,1}+a_{k,1})}
{(\lambda_{kag_k,0,1}-\lambda_{kag_k,0,2})} \, (\lambda_{kag_k,0,1})^m 
+ \frac{(a_{k,0}\lambda_{kag_k,0,2}+a_{k,1})}
{(\lambda_{kag_k,0,2}-\lambda_{kag_k,0,1})} \, (\lambda_{kag_k,0,2})^m \ . 
\label{pkagkfree}
\eeq
\end{widetext}
(Note that this is symmetric under the interchange $\lambda_{kag_k,0,1}
\leftrightarrow \lambda_{kag_k,0,2}$.)  For the relevant range of $q$,
$\lambda_{kag_k,0,1} > \lambda_{kag_k,0,2}$.  Therefore, in the limit $m \to
\infty$, the ground state degeneracy per vertex of this family of 
lattice strips is 
\beq
W(\{H_k(kag)\},q) = (\lambda_{kag_k,0,1})^{\frac{1}{5+2k}} \ , 
\label{wkagk}
\eeq
where the $\lambda_{kag_k,0,j}$ for $j=1,2$ were given in
Eq. (\ref{lamkagkd0}).  This and Eq. (\ref{pkagkfree}) are the main results of
the present paper. 

From the analytic result (\ref{wkagk}), there follow two monotonicity
properties: (i) for a given $k$, $W(\{H_k(kag)\},q)$ is a monotonically
increasing function of $q$ in the range $q \ge \chi=3$; and (ii) for a given $q
\ge 3$, $W(\{H_k(kag)\},q)$ is a monotonically increasing function of $k$ for
$k \ge 0$.  The fact that $W(\{G \},q)$ is an increasing function of $q$ for $q
\ge \chi(G)$ is quite general and is a consequence of the greater freedom in
performing proper $q$-colorings of $G$ for larger $q$.  Property (ii) can be
understood as a result of the fact that a proper $q$-coloring of a graph $G$
involves a constraint on the coloring of adjacent vertices of $G$, and this, in
turn, gives rise to a constraint from circuits in $G$.  Since the minimum
length of a circuit is the girth, increasing the girth tends to reduce the
severity of this latter constraint.  (Here, the girth of a graph $G$ is defined
as the number of edges that one traverses in a minimum-length circuit on
$G$.)  Although the girth of $[H_k(kag)]_{m,BC}$ (ignoring the double edges
that occur for $m=1$ with cyclic BC) is equal to 3, independent of $k$, the
girth of the polygons with $p=6+2k$ sides in the strip does increase with $k$.
Hence, for a fixed $q \ge \chi(G)=3$, this increase in the girth of the
$p$-gons increases the possibilities for proper $q$-colorings, and this, in
turn, increases the $W$ function.  These monotonicity properties are reflected
in the large-$q$ Taylor series expansions of the $W$ functions.  As $q \to
\infty$, the leading terms of the large-$q$ series expansions of
$q^{-1}W(\{H_k(kag)\},q)$ are of the form $q^{-1} W(\{H_k(kag) \},q) = 1 -
\alpha_k/q + ...$, where $...$ represents higher order terms in $1/q$, and the
coefficients for $k=0, \ 1, \ 2$ are $\alpha_0 = 8/5$, $\alpha_1 = 10/7$, and
$\alpha_2 = 4/3$, so that $\alpha_0 > \alpha_1 > \alpha_2$, and so forth for
higher $k$.

In Table \ref{wvalues} we list values of $W(\{kag\},q) \equiv
W(\{H_0(kag)\},q)$, $W(\{H_1(kag)\},q)$, and $W(\{H_2(kag)\},q)$ for
$3 \le q \le 10$.  The two general monotonicity properties stated above are
evident in the table.  As is also evident, $W(\{H_k(kag)\})$ is an
approximately linear function of $q$ for values of $q$ moderately above the
chromatic number, $\chi=3$.  

It is of interest to compare these results for the ground state degeneracy and
entropy on infinite-length limits of homeomorphic expansions of the kagom\'e
strip with those obtained for homeomorphic expansions of the square lattice
ladder strip in Ref. \cite{pg}.  The strip graphs considered in Ref. \cite{pg}
were constructed by starting with a free or cyclic (or M\"obius) square-ladder
strip of $m$ squares and adding $k-2$ vertices to each longitudinal edge, with
$k \ge 2$. Thus, the parameter $k-2$ of Ref. \cite{pg} corresponds to the
parameter $k$ in our present notation, and the resultant strip is (with our
present notational convention for $k$) $[H_k(sq)]_{m,BC}$.  This graph is thus
a homopolygonal strip of $p'$-gons, where $p'=2k+4$. We denote the $m \to
\infty$ limit of this strip as $\{[H_k(sq)]_{BC}\}$.  For $\{[H_k(sq)]_c \}$,
$q_c=2$ (independent of $k$) and, for $q \ge q_c$, $W$ is the same for the free
and cyclic (and M\"obius) longitudinal boundary conditions;
$W(\{[H_k(sq)]_f\},q) = W(\{[H_k(sq)]_c\},q) \equiv W(\{H_k(sq)\},q)$.
Converting the result of Ref. \cite{pg} to our present notation by the
replacement $k-2 \to k$, one has
\beq
W(\{H_k(sq)\},q) = (D_{2k+4})^{\frac{1}{2k+2}} \ . 
\label{wch}
\eeq
In general, for even $p'=2k+4$, (i) $D_{p'}=1$ if $q=2$ and hence 
$W(\{H_k(sq)\},2)=1$; (ii) $D_{p'}$ is a monotonically increasing function of
$q$, and hence so is $W(\{H_k(sq)\},q)$; (iii) for a given $q > 2$, 
$W(\{H_k(sq)\},q)$ is a monotonically increasing function of $k$. 
This monotonic increase as a function of the homeomorphic expansion parameter
$k$ is understandable in a manner analogous to that explained above, with the
difference that whereas the girth of the $[H_k(kag)]_{m,BC}$ strip is 3,
independent of $k$, the girth of $[H_k(sq)]_{m,BC}$ is $p'$. 

The comparison of the exact analytic result (\ref{wch}) for $W(\{
H_k(sq) \},q)$ from Ref. \cite{pg} for homeomorphic expansions of the
square-lattice ladder strip with our result (\ref{wkagk}) for homeomorphic
expansions of the kagom\'e strip yields another inequality, namely that for $q
\ge 3$ (so that one can perform a proper $q$-coloring of the
$[H_k(kag)]_{m,BC}$ strip),
\beq
W(\{H_k(kag)\},q) < W(\{H_k(sq)\},q) \ . 
\label{wineq}
\eeq
This inequality can be understood heuristically as follows. As before, it will
suffice to use free longitudinal boundary conditions and hence the graphs
$[H_k(sq)]_{m,f}$ and $[H_k(kag)]_{m,f}$ for the $m \to \infty$ limits that
define the respective $W$ functions.  Roughly speaking, for a given $k$, the
larger $q - \chi(G)$ is for a given graph $G$, the more freedom there is in
performing proper $q$-colorings of this graph. Now, for any $k$, the chromatic
number $\chi$ is larger (namely, 3) for $[H_k(kag)]_{m,f}$ than for
$[H_k(sq)]_{m,f}$ (namely, 2).  Hence, for $q$ greater than the larger of the
two chromatic numbers on these strips, $q-\chi(G)$ is larger for the
homeomorphic expansion of the square strip than for the homeomorphic expansion
of the kagom\'e strip.  The resultant greater freedom in performing proper
$q$-colorings of $[H_k(sq)]_{m,f}$ than of $[H_k(kag)]_{m,f}$ makes the
inequality (\ref{wineq}) understandable.

\section{Cyclic Strip $[H_k(kag)]_{m,c}$ } 

Using similar methods, we have calculated the chromatic polynomial for the 
homeomorphically expanded cyclic kagom\'e strip, $P([H_k(kag)]_{m,c},q)$. 
We find that (using the abbreviation $kag_k = [H_k(kag)]_c$ here) 
\beq
P([H_k(kag)]_{m,c},q)= \sum_{d=0}^2 c^{(d)} \sum_{j=1}^{n_P(kag_k,d)} \, 
(\lambda_{kag_k,d,j})^m
\label{pkagcyc}
\eeq
where $c^{(0)}=1$, $c^{(1)}=q-1$, and $c^{(2)}=q^2-3q+1$, and
\beq
n_P(kag_k,0) = 2, \quad n_P(kag_k,1) = 3, \quad n_P(kag_k,2) = 1 \ , 
\label{npkagkd}
\eeq
independent of $k$.  Hence, the total number of $\lambda$ terms that enter in  
Eq. (\ref{pkagcyc}) is 
\beq
N_{P,[H_k(kag)]_c,\lambda}=6 \ , 
\label{nptotkagkd}
\eeq
independent of $k$. Our structural result (\ref{pkagcyc}) showing the role that
the coefficients $c^{(d)}$ play for these homeomorphic expansions of a cyclic
kagom\'e strip graph generalizes what had been established earlier, namely that
they occur for the corresponding homeomorphic expansions of a square-lattice
strip graph \cite{pg} and for (non-homeomorphically expanded) cyclic strips of
the square \cite{saleur,cf}, triangular \cite{cf}, and honeycomb \cite{hca}
strip graphs, with the maximal $d$ corresponding to the width, $L_y$.  Although
it is not needed here, we recall the general formula
\beq
c^{(d)} = \sum_{s=0}^d (-1)^s {2d-s \choose s} q^{d-s} \ . 
\label{cd}
\eeq

We give the $\lambda$ terms that enter in Eq. (\ref{pkagcyc}) next. 
As is true in general for these recursive strip graphs \cite{s5}, the
$\lambda$'s that occur for the strip with free longitudinal boundary
conditions, $\lambda_{kag_k,0,j}$ (given in Eq. (\ref{lamkagkd0})), are
the same as the $\lambda$'s with $d=0$ in Eq. (\ref{pkagcyc}) for the cyclic
strip.  Note that 
\beq
\lambda_{kag_k,0,1}\lambda_{kag_k,0,2} = b_{k,2} = (q-1)^{3+2k}(q-2)^3 \ . 
\label{lamkagkd0product}
\eeq
At $q=0$, 
\beq
(\lambda_{kag_k,0,j})_{q=0} = -2(p-1 \pm \sqrt{p^2-2p-1} \ ) \ . 
\label{lamkagkd0q0}
\eeq
For the $\lambda$'s with $d=1$, we find, first,  
\beq
\lambda_{kag_k,1,1} = (-1)^k(q-1)^{1+k}(q-2)^2 \ . 
\label{lamkagkd1j1}
\eeq
Let us define
\beq
S_{k,1} = q-4+(-1)^k(q-2)(D_{k+4}-2D_{k+3}+D_{k+2})
\label{lamd1sum}
\eeq
and
\beq
P_{k,1} = (-1)^k(q-1)^{1+k}(q-2)^3 \ .
\label{lamd1prod}
\eeq
Then 
\beq \lambda_{kag_k,1,j} = \frac{1}{2}(S_{k,1} \pm \sqrt{S_{k,1}^2-4P_{k,1}} \
) \ ,
\label{lamkagd1j}
\eeq
where $j=2,3$ corresponds to the $\pm$ sign. Thus, 
\beq
\lambda_{kag_k,1,2}\lambda_{kag_k,1,3}=P_{k,1} 
\label{lamkagkd1j2j3product}
\eeq
so that
\beq
\prod_{j=1}^3 \lambda_{kag_k,1,j} = (q-1)^{2(1+k)}(q-2)^5 \ . 
\label{lamkagkd1j1j2j3product}
\eeq
For the $\lambda$ with $d=2$, we calculate 
\beq
\lambda_{kag_k,2} = q-4 \ . 
\label{lamkagd2}
\eeq
independent of $k$. It is easily checked that the $k=0$ special case of these
general results agrees with the calculation of the chromatic polynomial for the
cyclic kagom\'e strip in \cite{wcyl}.

\section{Locus ${\cal B}$}

From Eq. (\ref{fk}), it follows that $P(G,q)$ can be written in terms of its
zeros (called chromatic zeros) $q_{zj}$, $j=1,...,n$, as $P(G,q) =
\prod_{j=1}^n (q-q_{zj})$. These zeros are a natural topic for study in the
context of chromatic polynomials.  For a strip graph such as the ones
considered here, as $m \to \infty$, chromatic zeros merge to form an asymptotic
accumulation set (locus) consisting of various curves.  As in our earlier work,
we denote this locus as ${\cal B}$.  This locus is the solution to the equation
of degeneracy in magnitude of the dominant $\lambda$'s (i.e., the $\lambda$'s
with the largest absolute value in the complex $q$ plane \cite{bkw}).

\subsection{Case of Free Longtudinal Boundary Conditions}

For the $m \to \infty$ limit of the free strip $[H_k(kag)]_{m,f}$, the locus
${\cal B}$ involves a set of curves forming arcs.  For the kagom\'e strip
itself (i.e., the case $k=0$), these were shown in Fig. 7 of Ref. \cite{strip},
and we find a similar arc-like structure for $k \ge 1$.  The arc endpoints
occur at the zeros of the polynomial $R_{kkd0}$ given in Eq. (\ref{rkagkd0}). 
For example, for the actual kagom\'e strip itself, this is a polynomial of
degree 8, with zeros at
\beqs
q_1, \ q_1^* & = & 0.41 \pm 0.955i, \cr\cr
q_2, \ q_2^* & = & 1.18 \pm 1.14i, \cr\cr
q_3, \ q_3^* & = & 1.80 \pm 1.19i, \cr\cr
q_4, \ q_4^* & = & 2.62 \pm 0.15i  \ . 
\label{rk0d0zeros}
\eeqs
In this case ${\cal B}$ consists of four arcs, forming two complex-conjugate 
pairs, namely an arc connecting $q_1$ and $q_2$, an arc connecting $q_3$ and
$q_4$, and the complex-conjugate arcs. For general $k$, $R_{kkd0}$ is a
polynomial in $q$ of degree 
\beq
{\rm deg}(R_{kkd0}) = 8 + 4k \ . 
\label{degrkkd0}
\eeq
For this case of of $m \to \infty$ limit of $[H_k(kag)]_{m,f}$ with general 
$k$, the locus ${\cal B}$ consists of $4+2k$ arcs consisting of
$2+k$ complex-conjugate pairs, with endpoints at the $8 + 4k$ zeros of 
$R_{kkd0}$.

\subsection{Case of Cyclic Longtudinal Boundary Conditions}

The analysis of the locus ${\cal B}$ is more complicated for the $m \to \infty$
limit of the cyclic $[H_k(kag)]_{m,c}$ strips because of the presence of more
$\lambda$'s, namely six in all.  Again, the locus is determined by the equality
in magnitude of two dominant $\lambda$'s.  For the infinite-length limit of a
given family of graphs $\{G\}$, the maximal point at which ${\cal B}$ crosses
the real axis is denoted $q_c(\{G\})$.  As our previous work showed, for
families of graphs with free longitudinal boundary conditions, ${\cal B}$ does
not necessarily cross the real axis.  However, for families of graphs with
cyclic boundary conditions, ${\cal B}$ always crosses the real axis, so a $q_c$
is defined.  For the $m \to \infty$ limit of the $[H_k(kag)]_{m,c}$ graphs, 
considered here, denoted as $\{ [H_k(kag)]_c \}$, $q_c$ is determined by the 
equality of the dominant $\lambda$'s 
\beq
|\lambda_{kag_k,0,1}| = |\lambda_{kag_k,2}|=|q-4| \ . 
\label{qceq}
\eeq
For the infinite-length limit of the cyclic kagom\'e strip, 
$\{kag_c\}$ \cite{wcyl}, 
\beq
q_c(\{kag_c \}) \simeq 2.62 \ . 
\label{qckag}
\eeq
In the thermodynamic limit of the 2D kagom\'e lattice, previous work suggests
that $q_c(kag,2D)=3$ \cite{wurev}.  Hence, one sees that the $q_c$ value for
this kagom\'e strip is already within about 13 \% of the value for the infinite
2D lattice.  For the $[H_k(kag)]_{m,c}$ graphs, as $k$ increases, the effect of
the $p$-gons with $p=6+2k$ becomes greater, so one expects that $q_c$ will
decrease as $k$ increases, since $q_c=2$ for the $m \to \infty$ limit of the
circuit graph $C_m$.  Our exact results confirm this expectation.  For example,
for the infinite-length limits of the $[H_k(kag)]_{m,c}$ strips with $k=1$,
$k=2$, and $k=3$, we find
\beq
q_c(\{[H_1(kag)]_c\} \simeq 2.52 \ , 
\label{qckag1}
\eeq
\beq
q_c(\{[H_2(kag)]_c\}) \simeq 2.44 \ , 
\label{qckag2}
\eeq
and
\beq
q_c(\{[H_3(kag)]_c\}) \simeq 2.38 \ . 
\label{qckag3}
\eeq
The boundary ${\cal B}$ crosses the real $q$ axis at $q=0$, $q=2$, and $q=q_c$.
The degeneracy of $\lambda$ magnitudes at $q_c$ was given above in
Eq. (\ref{qceq}).  At $q=0$ there is a degeneracy in magnitude between
$\lambda_{kag_k,0,1}$ and the dominant $\lambda_{kag_k,1,j}$, $j=2,3$.  At
$q=2$, there is a degeneracy in magnitude between this dominant
$\lambda_{kag_k,1,j}$ and $\lambda_{kag_k,2}$, with both having magnitude equal
to 2.  There are thus three regions that include parts of the real axis. Region
$R_1$ includes the two semi-infinite line segments $q > q_c$ and $q < 0$ and
extends outward infinitely far from the origin.  In region $R_1$,
$\lambda_{kag_k,0,1}$ is the dominant $\lambda$ (i.e., the one with the largest
magnitude).  Region $R_2$ includes the interval $2 \le q \le q_c$.  In region
$R_2$, $\lambda_{kag_k,2}=q-4$ is the dominant $\lambda$.  Region $R_3$
includes the real interval $0 \le q \le 2$, and in this region, the dominant
term is the maximal-magnitude $\lambda_{kag_k,1,j}$ for $j=2,3$.  Other
complex-conjugate bubble phases are also present, as was found in
Ref. \cite{pg} and \cite{wcyl}.  Indeed, as is evident from Fig. 2 of
Ref. \cite{wcyl}, for the infinite-length strip of the cyclic kagom\'e lattice
itself, the boundary ${\cal B}$ encloses two very small complex conjugate
phases centered at approximately $q \simeq 2.53 \pm 0.50i$.

\section{Conclusions}

In conclusion, we have presented exact calculations of the chromatic polynomial
and ground state degeneracy and entropy per site of the $q$-state Potts
antiferromagnet on lattice strips that are homeomorphic expansions of a free or
cyclic strip of the kagom\'e lattice. These results show how $W$ and hence
$S_0$ increase as functions of the homeomorphic expansion parameter $k$.  We
have also compared the values of $W$ computed for the infinite-length limits of
these homeomorphically expanded kagom\'e strips with corresponding calculations
for kagom\'e strips without homeomorphic expansion given in
Refs. \cite{wcyl,wcy} and for homeomorphic expansions of square-lattice ladder
strips given in Ref. \cite{pg}.  Our present results yield further interesting
insights into the effect of homeomorphic graph expansions on nonzero ground
state entropy in the Potts antiferromagnet.

\bigskip

\begin{acknowledgments}

This research was partially supported by the grant NSF-PHY-06-53342.

\end{acknowledgments}

\begin{widetext}
\begin{table}
\caption{\footnotesize{Values of $W(\{kag\},q) \equiv W(\{H_0(kag)\},q)$,
$W(\{H_1(kag)\},q)$, and $W(\{H_2(kag)\},q)$ for $3 \le q \le 10$. For
comparison, we also show $W(\{sq \},q) \equiv W(\{H_0(sq)\},q)$,
$W(\{H_1(sq)\},q)$, and $W(\{H_2(sq)\},q)$ for the square-lattice ladder
strips. To save space, we omit the argument $q$ in these $W$ functions
below. See text for further details}}
\begin{center}
\begin{tabular}{ccccccc}
$q$ & $W(\{kag\})$ & $W(\{H_1(kag)\})$ &  $W(\{H_2(kag)\})$ 
& $W(\{sq\})$ & $W(\{H_1(sq)\})$ &  $W(\{H_2(sq)\})$ 
\\ \hline
 3 & 1.409  & 1.550  & 1.639  & 1.732  & 1.821 & 1.872 \\
 4 & 2.410  & 2.564  & 2.655  & 2.646  & 2.795 & 2.860 \\
 5 & 3.410  & 3.569  & 3.660  & 3.606  & 3.784 & 3.854 \\
 6 & 4.410  & 4.571  & 4.663  & 4.583  & 4.778 & 4.850 \\
 7 & 5.409  & 5.571  & 5.664  & 5.568  & 5.773 & 5.848 \\
 8 & 6.408  & 6.572  & 6.665  & 6.557  & 6.770 & 6.846 \\
 9 & 7.407  & 7.572  & 7.665  & 7.550  & 7.768 & 7.844 \\
10 & 8.407  & 8.572  & 8.665  & 8.544  & 8.766 & 8.843 \\
\hline
\end{tabular}
\end{center}
\label{wvalues}
\end{table}
\end{widetext}

\vfill
\eject

\begin{thebibliography}{99}

\bibitem{pauling35}
L. Pauling, J. Am. Chem. Soc. {\bf 57}, 2680 (1935).

\bibitem{berg07}
B. A. Berg, C. Muguruma, and Y. Okamoto, Phys. Rev. B {\bf 75}, 092202 (2007).

\bibitem{wurev}
F. Y. Wu, Rev. Mod. Phys. {\bf 54}, 235 (1982).

\bibitem{baxterbook}
R. J. Baxter, {\it Exactly Solved Models} (Oxford Univ. Press, Oxford, 1982).

\bibitem{chowwu}
Y. Chow and F. Y. Wu, Phys. Rev. B {\bf 36}, 285 (1987).

\bibitem{hs}
R. Shrock and S.-H. Tsai, Physica A {\bf 259}, 315 (1998).

\bibitem{pg}
R. Shrock and S.-H. Tsai, J. Phys. A Letts. {\bf 32}, L195-L200 (1999).

\bibitem{strip}
M. Ro\v{c}ek, R. Shrock, and S.-H. Tsai, Physica {\bf A252}, 505 (1998).

\bibitem{wcyl}
R. Shrock and S.-H. Tsai,  Phys. Rev. {\bf E60}, 3512 (1999).

\bibitem{wcy}
R. Shrock and S.-H. Tsai, Physica A {\bf 275}, 429 (2000).

\bibitem{biggs}
N. Biggs, {\it Algebraic Graph Theory} (Cambridge Univ. Press, Cambridge,
1993). 

\bibitem{dktbook}
Dong, F. M., Koh, K. M., Teo, K. L.: {\it Chromatic Polynomials and
Chromaticity of Graphs} (World Scientific, Singapore, 2005).

\bibitem{newton}
%
{\it Workshop on
Zeros of Graph Polynomials}, Newton Institute for Mathematical Sciences,
Cambridge University (2008), 
\newline http://www.newton.ac.uk/programmes/CSM/seminars.

\bibitem{w}
R. Shrock and S.-H. Tsai,  Phys. Rev. E {\bf 55}, 5165 (1997).

\bibitem{biggs77}
N. L. Biggs, Bull. London Math. Soc. {\bf 9}, 54 (1977).

\bibitem{w3}
R. Shrock and S.-H. Tsai, Phys. Rev. E {\bf 56}, 2733 (1997).

\bibitem{wn}
R. Shrock and S.-H. Tsai, Phys. Rev. E {\bf 56}, 4111 (1997).

\bibitem{wsc}
R. Shrock and Y. Xu, Phys. Rev. E {\bf 81}, 031134 (2010). 

\bibitem{wz84}
E. G. Whitehead and L. C. Zhao, J. Graph Theory {\bf 8}, 355 (1984).

\bibitem{wa3}
R. Shrock and S.-H. Tsai, J. Phys. A {\bf 31}, 9641 (1998).

\bibitem{wa2}
R. Shrock and S.-H. Tsai, Physica A {\bf 265}, 186 (1999).

\bibitem{rw99}
R. C. Read and E. G. Whitehead, Discrete Math. {\bf 204}, 337 (1999). 

\bibitem{sokal04}
A. Sokal, Combin. Probab. Comput. {\bf 13}, 221 (2004). 
  
\bibitem{peng04}
Y. L. Peng, Discrete Math. {\bf 288}, 177 (2004). 

\bibitem{fk}
C. M. Fortuin and P. W. Kasteleyn, Physica {\bf 57}, 536 (1972).

\bibitem{saleur}
H. Saleur, Nucl. Phys. B {\bf 360}, 219 (1991). 

\bibitem{cf}
S.-C. Chang and R. Shrock, Physica A {\bf 296}, 131 (2001).

\bibitem{hca}
S.-C. Chang and R. Shrock,  Physica A {\bf 296}, 183 (2001).

\bibitem{s5}
S.-C. Chang and R. Shrock, Physica A {\bf 316}, 335 (2002).

\bibitem{bkw}
S. Beraha, J. Kahane, and N. Weiss, J. Combin. Theory B {\bf 28}, 52 (1980).

\end{thebibliography}
\end{document}